\documentclass[pdflatex,sn-basic]{sn-jnl}% Basic Springer Nature Reference Style/Chemistry Reference Style

\usepackage{longtable,booktabs,amsmath,etoolbox,xcolor,natbib}   
\jyear{2022}%

\theoremstyle{thmstyleone}%
\theoremstyle{thmstyletwo}%

\theoremstyle{thmstylethree}%

\definecolor{brickred}{rgb}{0.8, 0.25, 0.33}

\begin{document}

\title[hermiter: R package for Sequential Nonparametric Estimation]{hermiter: R package for Sequential Nonparametric Estimation}

\author*[1]{\fnm{Michael} \sur{Stephanou}}\email{michael.stephanou@gmail.com}

\author[2,3]{\fnm{Melvin} \sur{Varughese}}\email{melvin.varughese@gmail.com}

\affil*[1]{\orgname{Rand Merchant Bank}, \orgaddress{\city{Johannesburg}, \country{South Africa}}}

\affil[2]{ \orgname{University of Cape Town}, \orgaddress{\city{Cape Town}, \country{South Africa}}}

\affil[3]{\orgname{University of Western Australia}, \orgaddress{\city{Perth}, \country{Australia}}}

\abstract{This article introduces the \textbf{R} package \textsf{hermiter} which facilitates estimation of univariate and bivariate probability density functions and cumulative distribution functions along with full quantile functions (univariate) and nonparametric correlation coefficients (bivariate) using Hermite series based estimators. The algorithms implemented in the \textsf{hermiter} package are particularly useful in the sequential setting (both stationary and non-stationary) and one-pass batch estimation setting for large data sets. In addition, the Hermite series based estimators are approximately mergeable allowing parallel and distributed estimation.}



%%\pacs[MSC Classification]{60-08}

\maketitle

\section{Introduction} \label{sec:intro}

Efficient algorithms for the nonparametric statistical analysis of streaming data and massive data sets are becoming increasingly relevant. Indeed, given the recent ubiquity of such data there is a growing need for such algorithms. Ideally these algorithms should be able to update estimates with constant i.e. $O(1)$ time and space complexity. This would allow effective processing of streaming data along with estimation on arbitrarily large data sets with fixed amounts of memory available. In addition, these algorithms should ideally be able to treat both stationary (static) and non-stationary (dynamic) settings, where statistical estimands are constant or may vary over time respectively. Finally, it would be very useful for these algorithms to also facilitate decentralized estimation where analyses on different subsets of a larger data set can be consistently and efficiently merged. Fundamental statistical quantities of interest for online nonparametric analysis include the mean, variance, higher order moments, probability density function (PDF), cumulative distribution function (CDF) and quantiles in the univariate setting. The most complete distributional information is provided by PDFs, CDFs and associated quantiles. The quantiles also have the favorable property of being robust to outliers and are more appropriate than moments such as the variance for skewed data. 

In the bivariate setting PDFs, CDFs and correlations are core quantities of interest. The most commonly applied measure of correlation is the Pearson product-moment correlation coefficient. This measure has several limitations however. The first is that it only captures the linear aspect of relationships between the random variables. The second is that the sample estimator for the Pearson product-moment correlation is not robust and is thus sensitive to outliers. Finally, this measure of correlation is not invariant under all order preserving transformations of the random variables. Nonparametric correlation measures such as the Spearman rank correlation coefficient (Spearman Rho) and the Kendall rank correlation coefficient (Kendall Tau) do not suffer from these limitations. They are appropriate to the broader class of monotonic relationships, are robust to outliers \citep{croux2010influence}, and are invariant under all order preserving transformations \citep{gibbons2010nonparametric}. 

In this article we present algorithms and software - i.e the R package \textsf{hermiter} - based on Hermite series estimators, that focus on the most general statistical quantities, namely the PDF, CDF and quantiles in the univariate setting and the PDF, CDF and nonparametric correlation coefficients in the bivariate setting. The algorithms we present have $O(1)$ time and space update complexity, can treat static and dynamic estimation and are fully mergeable (albeit only approximately in some cases). This mergeability implies suitability for parallelized and distributed computation of all the aforementioned estimands. Parallel computation could be carried out on a single machine, utilizing multiple cores/threads or split across several machines. Both situations allow faster computation. The fact that Hermite series based estimators are appropriate for distributed computation suggests another natural use case, namely edge computing. For example, different machines on a network could each update a Hermite series estimator with observations from attached sensors. Since the representation of the Hermite estimator is very compact, each estimator could then be efficiently communicated to a central data center to be merged into a single estimator that can provide combined metrics.

This article is organized as follows. In section \ref{sec:related_work} we review existing sequential algorithms and highlight some advantages and disadvantages of these algorithms. This is followed by a comparison of the existing algorithms with the Hermite series based algorithms in section \ref{sec:comparison_existing}. Related software is reviewed in section \ref{sec:related_software} after which we discuss our software contribution in section \ref{sec:our_package}. In section \ref{sec:est_algorithm} we review the Hermite series based estimators and algorithms that are implemented in \textsf{hermiter}. In section \ref{sec:merge_hermite_est} we treat merging of the estimators. In section \ref{sec:examples} we present concrete examples of code to construct a \texttt{hermite\_estimator} object, update it, merge it and calculate univariate and bivariate statistics. In section \ref{sec:numerical_example} we present a numerical example utilizing the \textsf{hermiter} package to analyze real foreign exchange data. In section \ref{sec:acc_perf} we briefly review existing accuracy comparisons between Hermite series based estimators and competing approaches. In addition, we present new results comparing Hermite series based algorithms as implemented in \textsf{hermiter} to the tdigest algorithm for online quantile estimation in the univariate setting and to the algorithm in \cite{xiao2019novel} for online estimation of the Kendall Tau coefficient in the bivariate setting. We conclude in section \ref{sec:summary}.

\section{Related work}\label{sec:related_work}

\subsection{Sequential PDF, CDF and quantile estimation}

There are existing algorithms for online estimation of all of the aforementioned statistical quantities. For PDFs, existing algorithms include recursive kernel estimators as discussed in Chapters 4 and 5 of \cite{greblicki2008nonparametric} and Chapter 7 of \cite{devroye1985nonparametric}. See also \cite{slaoui2019recursive} for recursive estimators based on Bernstein polynomials. In the case of CDFs, existing algorithms include those based on the empirical distribution function estimator, the smooth kernel distribution function estimator \citep{slaoui2014stochastic} and Bernstein polynomials \citep{jmaei2017recursive}. These PDF and CDF estimators furnish sequential estimates of the probability density and cumulative probability at predefined values of the support of the probability density function, which is not as general as the online estimation of the full PDF and CDF. 

Several algorithms exist for estimating quantiles in a sequential (one-pass) manner. These algorithms can be differentiated into two broad categories. The first category is comprised of those algorithms that directly approximate arbitrary empirical quantiles of a set of numeric values using sub-linear memory. Here, the empirical p-quantile is defined as $x_{(\lceil pn \rceil)}$ for $n$ observations, where $x_{(i)}, \, i \in 1,\dots n$, are the observations sorted in ascending order. No particular data generating process is presupposed. These algorithms provide either deterministic or probabilistic guarantees on the rank error of approximate sample quantiles and have space requirements that grow with required accuracy and potentially also with the number of observations. Noteworthy algorithms in this category include those due to \cite{greenwald2001space} and \cite{karnin2016optimal}. For recent reviews of algorithms of this type see \cite{greenwald2016quantiles}, \cite{luo2016quantiles} and \cite{chen2020survey}. Rank error accuracy is but one choice of quality metric however and it has been noted that controlling rank error accuracy does not preclude arbitrarily large relative error (for heavily skewed, or heavy tailed data) as per \cite{masson2019ddsketch}. In \cite{masson2019ddsketch} an algorithm providing relative error accuracy guarantees for estimating sample quantiles (with space requirements that grow with accuracy) is proposed to address this (see also \cite{epicoco2020uddsketch} for an enhancement of this algorithm). It is also worth noting that many of the aforementioned algorithms cannot be used to form dynamic quantile estimates based on a recent subset of observations since they are insertion-only algorithms. Here insertion-only means that the estimates can only be updated with new observations in an online manner, but previous observations cannot be removed in such a manner. Exceptions include those algorithms that also allow deletion operations and use turnstile semantics to maintain an estimate over a sliding window. These algorithms include the Dyadic Count-Sketch \citep{luo2016quantiles} for example. Such algorithms assume a fixed universe of potential values for the values being analyzed and typically have higher space and update time complexity than insertion-only algorithms. See also \cite{zhao2021kll} for recent developments using a bounded deletion model.  

The second broad category of online quantile estimation algorithms, of which the Hermite series based algorithm is a member (i.e. the algorithm proposed in \cite{stephanou2017sequential} and implemented in \textsf{hermiter}), assume that the observations to be analyzed are a realization of a random process. Approximating quantiles is then a nonparametric statistical estimation problem. Algorithms in this category are numerous \citep{P2, extP2First,extP2Second, P2ewma,ewsa,yazidi2017multiplicative, hammer2019tracking, hammer2019new, tiwari2019technique}. The aforementioned algorithms in the literature only apply to pre-specified quantiles however. A recently introduced algorithm by \cite{gan2018moment} involving moment-based quantile sketches, furnishes an efficient approach to quantile estimation but focusses on rapid mergeability and single quantile queries. See also \cite{mitchell2021empirical} for a comparison of the moment-based quantile sketch algorithm with several other algorithms on a large collection of real-world data sets. In \cite{mitchell2021empirical} orthogonal series based quantile estimators similar to those in \cite{stephanou2017sequential} are studied, including an estimator based on the Chebyshev-Hermite polynomials (also known as the ``Probabilist's" Hermite polynomials). These are different orthogonal polynomials than those used in \cite{stephanou2017sequential}, which are instead the ``Physicist's" Hermite polynomials, and thus the results in \cite{mitchell2021empirical} do not directly apply to the estimators implemented in \textsf{hermiter}. The popular t-digest approach \citep{dunning2021t} maintains an estimate of the empirical distribution function and allows the estimation of arbitrary quantiles, presenting similar capabilities to the Hermite series based approach. This algorithm has the useful property of maintaining an accuracy which is nearly constant relative to $p(1-p)$ for the p-quantile. In addition, the t-digest algorithm has many desirable properties in practice, such as approximate mergeability, and is fast and accurate.

\subsection{Sequential nonparametric correlation estimation}

In the bivariate setting, there is only one alternate approach to online estimation of nonparametric correlation coefficients known to the authors, namely the algorithms introduced in \cite{xiao2019novel} which involve discretizing the joint distribution of the random variables. 

\section{Comparison with existing algorithms} \label{sec:comparison_existing}

The algorithms we implement based on Hermite series estimators have certain advantages over the existing approaches discussed above. Firstly, the Hermite series based approach allows the online estimation of the full PDF and CDF. Existing algorithms only furnish online estimates of the probability density and cumulative probability at predefined values of the support of the probability density function. This makes the Hermite series algorithms more general.

Secondly, the Hermite series based algorithms present several advantages in the online quantile estimation setting.  In comparison to algorithms that directly approximate arbitrary empirical quantiles, the Hermite series based estimators introduced in \cite{stephanou2017sequential} have fixed i.e. $O(1)$ space requirements, but do not provide explicit guarantees on rank error accuracy in estimating sample quantiles. As noted in the previous section, the rank error accuracy metric does have shortcomings however. In addition, the Hermite series based algorithms facilitate online quantile estimation in non-stationary settings with small and constant i.e. O(1) memory requirements, a distinct advantage over the aforementioned turnstile algorithms.

When comparing to the existing online nonparametric statistical estimators, the Hermite series based approach allows the online estimation of arbitrary quantiles at any point in time (along with a direct online estimate of the full CDF) compared to pre-specified quantiles only. The t-digest approach provides the most similar capabilities to the Hermite series based approach. As we will demonstrate though, the Hermite series based methods appear to have superior accuracy in the cases we studied with roughly comparable speed. In addition, the Hermite series based algorithms facilitate online quantile estimation in non-stationary settings with small and constant memory requirements whereas the t-digest algorithm does not allow dynamic quantile estimation directly. This is pertinent for tracking the quantile function of non-stationary processes in near real-time for example. It should be noted that a similar effect can be achieved by maintaining a fixed number of t-digest sketches for recent non-overlapping windows of data (including the most recent, potentially partially complete window) and merging these sketches prior to quantile estimation. Sketches for windows beyond a certain age could be excluded from being merged or deleted. This is perhaps best suited to maintaining continuously updated aggregates in the database context however\footnote{We would like to thank Ted Dunning for useful discussions in this regard.}. 

In the bivariate setting, the Hermite series based algorithms introduced in \cite{stephanou2021sequential} and in this article have accuracy advantages for large numbers of observations in comparison to the existing algorithms in \cite{xiao2019novel}. Our algorithms also allow the Spearman Rho and Kendall Tau of non-stationary streams to be tracked via an exponential weighting scheme, without relying on a moving/sliding window approach. This is advantageous since memory requirements are fixed with the exponential weighting approach compared to growing memory requirements with moving window size when utilizing a moving/sliding window approach.

\section{Review of related software} \label{sec:related_software}

Various open-source software implementations for online statistical estimation exist. Perhaps the most comprehensive solution is the \textsf{OnlineStats} package  \citep{day2020onlinestats} in \textbf{Julia} \citep{Julia-2017} which facilitates one-pass batch estimation, sequential estimation and merging of estimates of most of the aforementioned statistical quantities. Notable exceptions are bivariate PDFs, CDFs and nonparametric correlations. In \textbf{Python} \citep{pythonlang} the \textsf{RunStats} package \citep{runstatsPypkg} has a smaller coverage of statistical quantities than \textsf{OnlineStats} but provides similar functionality. Focussing more specifically on rapid online estimation of quantiles and cumulative probabilities, the tdigest algorithm has been implemented in several languages including \textbf{R} \citep{rcore} in the package \textsf{tdigest} \citep{tdigestRpkg}. A summary of the nonparametric statistics covered by each package is presented in Table \ref{tab:overview_packages}. Note that the package \textsf{OnlineStats} also facilitates online parametric estimation, which is not the focus of this article.  

\section{The R package \textsf{hermiter}} \label{sec:our_package}

The \textbf{R} package \textsf{hermiter} \citep{hermiterRpkg} which we introduce in this article, is a focussed package concentrating on online nonparametric estimation of specific quantities in the univariate and bivariate settings. It addresses a gap in current implementations, namely online estimation of bivariate probability density functions, cumulative distribution functions and nonparametric correlations. In addition, this package presents advantages over existing packages in certain univariate estimation problems. In particular, the \textsf{hermiter} package presents accuracy advantages in quantile estimation over a leading alternative \textbf{R} package, namely \textsf{tdigest} whilst having roughly comparable speed. The \textsf{hermiter} package can also directly handle non-stationary quantile estimation, which \textsf{tdigest} cannot. The unifying theme of the \textsf{hermiter} package is the use of recently introduced Hermite series based estimators for univariate and bivariate estimation. It collects the associated algorithms in a coherent implementation.

\begin{table}[t!]
\centering
\caption{\label{tab:overview_packages} Overview of open-source software implementations for online statistical estimation.}
{\footnotesize
\begin{tabular}{lllp{4cm}}
\hline
Package           & Language & Univariate Statistics   & Bivariate Statistics \\ \hline
\textsf{OnlineStats}            & \textbf{Julia}      & Mean, Variance, Quantiles, Extrema, \\
& & Skewness, Kurtosis, PDF, CDF       & Pearson's Correlation \\
\textsf{RunStats} & \textbf{Python}  & Mean, Variance, Extrema, Skewness,\\
& & Kurtosis & Pearson's Correlation \\
\textsf{tdigest}               & \textbf{R}              & CDF, Quantiles & - \\ 
\textsf{hermiter} & \textbf{R}      & Quantiles, PDF, CDF       & PDF, CDF, Spearman \\
& & & Rho, Kendall Tau\\ \hline
\end{tabular}
}
\end{table}

\section{Hermite series based estimators and algorithms} \label{sec:est_algorithm}

\subsection{Density estimation}

The Hermite series density estimator for the univariate PDF, $f(x)$, is defined as follows \citep{schwartz1967estimation,convergence1,convergence2, greblicki1985pointwise,liebscher1990hermite},
\begin{equation}
	\hat{f}_{N,n}(x) = \sum_{k=0}^{N} \hat{a}_{k}^{(n)} h_{k}(x), \quad \hat{a}_{k}^{(n)} = \frac{1}{n}\sum_{i=1}^{n} h_{k}(\mathbf{x_{i}}), \quad k = 0,\dots,N, \label{eq:HermiteSeriesProbEst}
\end{equation}
where $\mathbf{x_{i}} \sim f(x), \, i=1,2,\dots n$, and $h_{k}(x) =\left(2^{k}k!\sqrt{\pi}\right)^{-\frac{1}{2}} e^{-\frac{x^2}{2}} H_{k}(x), \, k=0,1,\cdots,N$ are the normalized Hermite functions defined from the Hermite polynomials,
$H_k(x)=(-1)^k e^{x^2}\frac{\mathrm{d}^k}{\mathrm{d}x^k}e^{-x^2},\, k=0,1,\cdots,N$. 

The bivariate Hermite series density estimator for the bivariate PDF, $f(x,y)$, has the following form:
\begin{align}
	&\hat{f}_{N_{1},N_{2},n}(x,y) = \sum_{k=0}^{N_{1}} \sum_{j=0}^{N_{2}} \hat{A}_{kj}^{(n)} h_{k}(x)h_{j}(y), \nonumber\\ 
	&\hat{A}_{kj}^{(n)} = \frac{1}{n}\sum_{i=1}^{n} h_{k}(\mathbf{x_{i}})h_{j}(\mathbf{y_{i}}). \quad k = 0,\dots,N_{1}, \, j=0,\dots,N_{2}, \label{eq:HermiteSeriesProbEstBivariate}
\end{align}
where $(\mathbf{x_{i}}, \mathbf{y_{i}}) \sim f(x,y), \, i=1,2,\dots n$. 
Note that although standardization of the observations is not required for Hermite series based estimators, our empirical studies have revealed that it often improves accuracy. Here standardization refers to subtracting the estimated mean from each observation and dividing by the estimated standard deviation. The estimates for the Hermite series coefficients based on these standardized observations are biased compared to the coefficients that would have been obtained by using the true mean and standard deviation for standardization. Despite this bias, we have typically observed better overall results in practice using standardization. The PDF estimators previously defined need to be modified for standardized observations. For the univariate case,
\begin{equation*}
	\hat{f}_{N,n}(x) = \frac{1}{\hat{\sigma}} \sum_{k=0}^{N} \hat{a}_{k}^{(n)} h_{k}(x),
\end{equation*}
where $\hat{\sigma}$ is the estimated standard deviation. For the bivariate case,
\begin{equation*}
	\hat{f}_{N_{1},N_{2},n}(x,y) = \frac{1}{\hat{\sigma}^{(1)}\hat{\sigma}^{(2)} } \sum_{k=0}^{N_{1}} \sum_{j=0}^{N_{2}} \hat{A}_{kj}^{(n)} h_{k}(x)h_{j}(y),
\end{equation*}
where $\hat{\sigma}^{(1)}$ and $\hat{\sigma}^{(2)}$ are the estimates of the standard deviation of $x$ and $y$ respectively.

\subsection{Distribution function and quantile estimation}

The univariate density estimator \eqref{eq:HermiteSeriesProbEst} can be used to define the following univariate CDF estimator,
\begin{equation}
	\hat{F}^{(1)}_{N,n}(x) = \sum_{k=0}^{N} \hat{a}_{k}^{(n)} \int_{-\infty}^{x} h_{k}(t) dt.\label{eq:HermiteSeriesUniCDFEst}
\end{equation}
Quantiles can be obtained numerically by applying a root-finding algorithm to solve for the quantile value, $\hat{x}_{p}$, at cumulative probability $p$ via,
\begin{equation*}
	\hat{F}^{(1)}_{N,n}(\hat{x}_{p}) = p.
\end{equation*}
In \cite{stephanou2017sequential} an alternate univariate CDF estimator was also introduced,
\begin{equation}
\hat{F}^{(2)}_{N,n}(x)\!=\!\begin{cases}
	 1 - \sum_{k=0}^{N} \hat{a}_{k}^{(n)} \int_{x}^{\infty} h_{k}(t) dt & \quad \mbox{if } x \geq 0, \\
	\sum_{k=0}^{N} \hat{a}_{k}^{(n)} \int_{-\infty}^{x} h_{k}(t) dt &  \quad \mbox{if } x < 0.
	\end{cases}
	\label{eq:HermiteSeriesUniCDFEstAlt}
\end{equation}
We have found that while the asymptotic properties of the alternate univariate CDF estimator \eqref{eq:HermiteSeriesUniCDFEstAlt} are the same as \eqref{eq:HermiteSeriesUniCDFEst}, the empirical, finite sample performance of \eqref{eq:HermiteSeriesUniCDFEstAlt} is superior for quantile estimation specifically, see \cite{stephanou2017sequential}. 
A vectorized implementation of the bisection root finding algorithm is a robust way to calculate the quantiles at an arbitrary vector of cumulative probability values. This is one of the two algorithms available for quantile estimation in \textsf{hermiter}. The second algorithm allows for more rapid but sometimes less accurate approximation of the quantiles by calculating the CDF values at a fixed vector of values and applying linear interpolation to find the vector of quantiles using the \textbf{R} function \textit{findInterval}. It is also noteworthy that in \eqref{eq:HermiteSeriesUniCDFEst} and \eqref{eq:HermiteSeriesUniCDFEstAlt}, it is computationally advantageous to utilize the recurrence relations for integrals of Hermite functions in \eqref{eq:intRecurrenceLower} and \eqref{eq:intRecurrenceUpper},
\begin{align}
	&\int_{-\infty}^{x} h_{0} (t) dt = \frac{\pi^{1/4}}{\sqrt{2}} \mbox{erfc}\left(\frac{-1}{\sqrt{2}} x\right),  \quad
	\int_{-\infty}^{x} h_{1} (t) dt =  -\frac{\sqrt{2}}{\pi^{1/4}}e^{-x^{2}/2},\nonumber\\
	&\int_{-\infty}^{x} h_{k+1} (t) dt = -\sqrt{\frac{2}{k+1}} h_{k} (x) 
	+ \sqrt{\frac{k}{k+1}} \int_{-\infty}^{x} h_{k-1} (t) dt, \, k=1,\dots, N, \label{eq:intRecurrenceLower}	
\end{align}
and
\begin{align}
	&\int_{x}^{\infty} h_{0} (t)  dt = \frac{\pi^{1/4}}{\sqrt{2}} \mbox{erfc}\left(\frac{1}{\sqrt{2}} x\right), \quad
	\int_{x}^{\infty}  h_{1} (t) dt = \frac{\sqrt{2}}{\pi^{1/4}} e^{-x^{2}/2},\nonumber\\
	&\int_{x}^{\infty}  h_{k+1} (t) dt = \sqrt{\frac{2}{k+1}} h_{k} (x)
	+ \sqrt{\frac{k}{k+1}} \int_{x}^{\infty} h_{k-1} (t) dt, \, k=1,\dots, N,\label{eq:intRecurrenceUpper}
\end{align}
where $\mbox{erfc}(x) = \frac{2}{\sqrt{\pi}} \int_{x}^{\infty} e^{-t^{2}}dt$ is the complementary error function. Convergence acceleration techniques can be successfully applied to improve accuracy at a given order of truncation for Hermite series. In particular, \textsf{hermiter} applies straightforward iterated averaging techniques to the truncated series (see \cite{boyd1986summability} and Table A.2 of \cite{boyd2018dynamics}) to improve the accuracy of univariate PDF and CDF estimates along with quantile estimates obtained through the CDF estimates. We have observed that empirically, applying series acceleration typically yields better results. Deriving formal guarantees that these methods are generally effective in our context is an area for future research. It is worth noting that the results presented in \cite{stephanou2017sequential} can be reproduced by setting the algorithm used for quantile estimation to \texttt{`bisection'} and the accelerate series argument to \texttt{FALSE}.
 
In a similar manner to the univariate case, the bivariate density estimator \eqref{eq:HermiteSeriesProbEstBivariate} can be used to construct the following bivariate CDF estimator:
 \begin{equation}
	\hat{F}_{N_{1},N_{2},n}(x,y) = \sum_{k=0}^{N_{1}} \sum_{j=0}^{N_{2}} \hat{A}_{kj}^{(n)} \int_{-\infty}^{x} h_{k}(t)dt \int_{-\infty}^{y} h_{j}(u) du. \label{eq:HermiteSeriesCDFEstBivariate}
\end{equation}
The recurrence relation in \eqref{eq:intRecurrenceLower} can again be applied for rapid calculation of the integrals.

\subsubsection{Limitations of Hermite series estimators}\label{sec:limitations}

The cumulative probability estimates produced by the estimators in this section may not be monotonically non-decreasing and may lie outside the range $[0,1]$. This is related to the fact that the Hermite series based distribution function estimators are obtained by integrating Hermite series density estimators (with a finite number of terms). These density estimators can, in principle, produce negative probability density estimates at certain values of the domain. In practice, these departures are usually quite minor given sufficient data and can be further ameliorated by clipping the cumulative probability estimates to lie between 0 and 1.

\subsection{Nonparametric correlation estimation}

We can also use the Hermite series based estimators defined above to obtain estimators for the Spearman Rho and Kendall Tau nonparametric correlation measures. For the Spearman correlation coefficient estimator, $\hat{R}_{N}$, we plug the bivariate Hermite PDF estimator and univariate Hermite CDF estimator into a large sample form of the Spearman correlation coefficient (see \cite{stephanou2021sequential}) to obtain, 
\begin{equation*}
     \hat{R}_{N}= 12\int\int(\hat{F}_{N, n}^{x}(x) - 1/2)(\hat{F}_{N, n}^{y}(y) - 1/2)\hat{f}_{N, N, n}(x,y) dx dy, 
\end{equation*}
where $\hat{F}_{N,n}^{x}(x),\hat{F}_{N,n}^{y}(y)$ correspond to  univariate CDF estimators as defined in equation   \eqref{eq:HermiteSeriesUniCDFEst}, updated on $x$ and $y$ observations respectively. The function $\hat{f}_{N, N, n}(x,y)$ is the bivariate PDF estimator defined in equation \eqref{eq:HermiteSeriesProbEstBivariate}. We can phrase the estimator above in terms of linear algebra operations for computational efficiency:
\begin{align}
     \hat{R}_{N}&= 12 \mathbf{\hat{a}_{(1)}}^{(n)}  \mathbf{W}^{T}\mathbf{\hat{A}}^{(n)}\mathbf{W}\mathbf{\hat{a}_{(2)}}^{(n)} 
- 6 \mathbf{\hat{a}_{(1)}}^{(n)} \mathbf{W}^{T} \mathbf{\hat{A}}^{(n)} \mathbf{z}
- 6 \mathbf{z} \mathbf{\hat{A}}^{(n)} \mathbf{W}\mathbf{\hat{a}_{(2)}}^{(n)} \nonumber\\
&+ 3  \mathbf{z} \mathbf{\hat{A}}^{(n)} \mathbf{z},\label{eq:hermiteSpearmansEstMat} 
\end{align}
where $\mathbf{\hat{a}_{(1)}}^{(n)}$ are the coefficients $(\mathbf{\hat{a}_{(1)}}^{(n)})_{k}$ associated with the Hermite CDF estimator $\hat{F}_{N,n}^{x}(x)$, $\mathbf{\hat{a}_{(2)}}^{(n)}$ are the coefficients $(\mathbf{\hat{a}_{(2)}}^{(n)})_{k}$ associated with the Hermite CDF estimator $\hat{F}_{N, n}^{y}(y)$ and $\mathbf{\hat{A}}^{(n)}$ are the coefficients $(\mathbf{\hat{A}}^{(n)})_{kl}$ associated with the Hermite bivariate PDF estimator $\hat{f}_{N,N,n}(x,y)$.  The matrix $(\mathbf{W})_{kl} = \int_{-\infty}^{\infty}h_{k}(u)  \int_{-\infty}^{u} h_{l}(v) dv du$ and vector $(\mathbf{z})_{k} = \int_{-\infty}^{\infty} h_k(u) du$. In all the  matrices and vectors we have defined,  $k,l = 0, \dots, N$. These integrals can be evaluated numerically once and the values stored for rapid calculation. The sequential nature of the estimators follows from the fact that we have obtained sequential update rules for the coefficients and the fact that there is no explicit dependence on the observations in these expressions as discussed in section \ref{subsec:seq_hermite_coeff} below. 

For the Kendall Tau estimator, $\hat{\tau}_{N}$, we plug the bivariate Hermite PDF estimator and bivariate Hermite CDF estimator into a form of the Kendall correlation coefficient relating directly to the probability of concordance (see \cite{gibbons2010nonparametric} for more details on this representation), to obtain,
\begin{equation*}
     \hat{\tau}_{N}= 4\left[\int\int \hat{F}_{N,N,n}(x,y)\hat{f}_{N,N,n}(x,y) dx dy\right] - 1,
\end{equation*}
where $\hat{F}_{N,N,n}(x,y)$ is the bivariate CDF estimator defined in equation \eqref{eq:HermiteSeriesCDFEstBivariate} and $\hat{f}_{N,N,n}$ is the bivariate PDF estimator defined in equation \eqref{eq:HermiteSeriesProbEstBivariate}. We can again phrase the estimator above in terms of linear algebra operations for computational efficiency:
\begin{equation}
     \hat{\tau}_{N} = 4 \mathbf{W}^{T}\left(\mathbf{\hat{A}}^{(n)}\right)^{T} \mathbf{W} \mathbf{\hat{A}}^{(n)}.\label{eq:hermiteKendallEstMat} 
\end{equation}

\subsection{Sequential Hermite series based estimators}\label{subsec:seq_hermite_coeff}

The particular utility of all the Hermite series based estimators in online/sequential estimation follows from the ability to update the estimates of the Hermite series coefficients in an online, $O(1)$ manner. In particular, for the coefficients in \eqref{eq:HermiteSeriesProbEst}, we can define the following updating algorithm,
\begin{equation}
                     \mathbf{\hat{a}^{(1)}} =   \mathbf{h} (\mathbf{x}_{1}), \quad
		\mathbf{\hat{a}^{(i)}} = \frac{1}{i} \left[(i-1)\mathbf{\hat{a}^{(i-1)}} + \mathbf{h} (\mathbf{x}_{i})\right], \, i=2,\dots,n. \label{eq:univariateUpdate}
\end{equation}
For the coefficients in \eqref{eq:HermiteSeriesProbEstBivariate}, we can define an updating algorithm in terms of the outer product of $\mathbf{h} (\mathbf{x})$ and $\mathbf{h} (\mathbf{y})$,
\begin{equation}
		\mathbf{\hat{A}}^{(1)} = \mathbf{h} (\mathbf{x}_{1}) \otimes \mathbf{h} (\mathbf{y}_{1}), \quad
		\mathbf{\hat{A}}^{(i)} = \frac{1}{i} \left[(i-1)\mathbf{\hat{A}}^{(i-1)} +\mathbf{h} (\mathbf{x}_{i}) \otimes \mathbf{h} (\mathbf{y}_{i})\right], \, i=2,\dots,n. \label{eq:bivariateUpdate}
\end{equation}
In the update rules \eqref{eq:univariateUpdate} and \eqref{eq:bivariateUpdate} above, $\mathbf{h} (\mathbf{x}_{i})$ and $\mathbf{h} (\mathbf{y}_{i})$ are the vectors $h_{k} (\mathbf{x}_{i}), \, k=0,\dots,N$ and $h_{l} (\mathbf{y}_{i}), l=0,\dots,N$ respectively. For computational efficiency it is advantageous to calculate $ \mathbf{h} (\mathbf{x}_{i})$ and $\mathbf{h} (\mathbf{y}_{i})$ making use of the recurrence relation for the Hermite polynomials $H_{k+1}(x) = 2xH_{k}(x) - 2kH_{k-1}(x)$. The computational cost of updating the coefficients, $\mathbf{\hat{a}}^{(i)}$ and $\mathbf{\hat{A}}^{(i)}$ as above is manifestly constant ($O(1)$) with respect to the number of previous observations. In non-stationary scenarios the following exponential weighting scheme can be applied for the coefficients defined in \eqref{eq:HermiteSeriesProbEst},
\begin{equation}
		\mathbf{\hat{a}^{(1)}} =   \mathbf{h} (\mathbf{x}_{1}), \quad
		\mathbf{\hat{a}^{(i)}} =  (1-\lambda) \mathbf{\hat{a}}^{(i-1)} + \lambda \mathbf{h}  (\mathbf{x}_{i}),  \, i=2,\dots,n,
		\label{eq:univariateUpdateExp}
\end{equation}
where $0 < \lambda \leq 1$ controls the weight of new terms associated with new observations. For the coefficients defined in \eqref{eq:HermiteSeriesProbEstBivariate} we have,	
\begin{equation}
		\mathbf{\hat{A}}^{(1)} = \mathbf{h} (\mathbf{x}_{1}) \otimes \mathbf{h} (\mathbf{y}_{1}), \quad
		\mathbf{\hat{A}}^{(i)} = (1-\lambda)\mathbf{\hat{A}}^{(i-1)} + \lambda \mathbf{h}  (\mathbf{x}_{i}) \otimes \mathbf{h} (\mathbf{y}_{i}), \, i=2,\dots,n.
		\label{eq:bivariateUpdateExp}
\end{equation}
The computational cost of updating the coefficients, $\mathbf{\hat{a}}^{(i)}$ and $\mathbf{\hat{A}}^{(i)}$ is again manifestly constant ($O(1)$) with respect to the number of previous observations.

As noted above, the Hermite series based estimators have been found to typically perform better on standardized observations. The observations can be standardized in an online manner by subtracting an online estimate of the mean and dividing by an online estimate of the standard deviation.

\section{Merging Hermite series based estimators}\label{sec:merge_hermite_est}

Hermite series based estimators can be consistently merged in both the univariate and bivariate settings. In particular, when the observations are not standardized, the results obtained from merging distinct Hermite series based estimators updated on subsets of a data set are exactly equal to those obtained by constructing a single Hermite series based estimator and updating on the full data set (corresponding to the concatenation of the aforementioned subsets). This follows from the fact that the merged coefficients of $m$ subsets of an overall data set,
\begin{align}
		(\mathbf{\hat{a}})_{\mbox{merged}} &= \sum_{j=1}^{m} \frac{n_{j}}{n} (\mathbf{\hat{a}})_{j}, \, \quad n = \sum_{j=1}^{m}  n_{j}, \label{eq:univariatMergeNonStd}
\end{align}
\begin{align}
		(\mathbf{\hat{A}})_{\mbox{merged}} &= \sum_{j=1}^{m} \frac{n_{j}}{n} (\mathbf{\hat{A}})_{j}, \, \quad n = \sum_{j=1}^{m}  n_{j}, \label{eq:bivariateMergeNonStd}
\end{align}
are exactly equal to the coefficients that would have been obtained on the full data set. Since the coefficient values are identical, it is easy to see that the PDF, CDF and quantile results in the univariate case and the PDF, CDF and nonparametric  correlation results in the bivariate case are identical too. When the observations are standardized, the equivalence is no longer exact but it is accurate enough to be practically useful. The algorithm we use is as follows, first the mean and standard deviation estimates from each of the estimators constructed on the data subsets are merged using the method in \cite{schubert2018numerically}. This procedure yields a merged mean, $\hat{\mu}_{\mbox{merged}}$, and a merged standard deviation, $\hat{\sigma}_{\mbox{merged}}$, that are exactly equal to the mean and standard deviation of the full data set. In the univariate case, we calculate the merged Hermite series coefficients as follows:
\begin{align}
		(\hat{a}_{k})_{\mbox{merged}} &= \sum_{j=1}^{m} \frac{n_{j}}{n} \int_{-\infty}^{\infty} h_{k}\left(\frac{\hat{\sigma}_{j} x + \hat{\mu}_j - \hat{\mu}_{\mbox{merged}}}{\hat{\sigma}_{\mbox{merged}}}\right) dx, \, \quad n = \sum_{j=1}^{m}  n_{j}. \label{eq:univariatMergeStd}
\end{align}
In the bivariate case, we utilize:
\begin{align}
		(\hat{A}_{kl})_{\mbox{merged}} &= \sum_{j=1}^{m} \frac{n_{j}}{n} \int_{-\infty}^{\infty}\int_{-\infty}^{\infty} h_{k}\left(\frac{\hat{\sigma}^{(1)}_{j} x + \hat{\mu}^{(1)}_j - \hat{\mu}^{(1)}_{\mbox{merged}}}{\hat{\sigma}^{(1)}_{\mbox{merged}}}\right) \times \nonumber \\
		&\times h_{l}\left(\frac{\hat{\sigma}^{(2)}_{j} y + \hat{\mu}^{(2)}_j - \hat{\mu}^{(2)}_{\mbox{merged}}}{\hat{\sigma}^{(2)}_{\mbox{merged}}}\right) dxdy, \, \quad n = \sum_{j=1}^{m}  n_{j}. \label{eq:bivariatMergeStd}
\end{align}
We use Gauss-Hermite quadrature to rapidly evaluate these integrals in the \textsf{hermiter} package. We conclude this section by contrasting the operation of sequential updating of a Hermite series based estimator with the operation of merging several estimators. In the case where  observations are not standardized, the result of repeated sequential updates to a single estimator versus repeated merging of several individual Hermite series based estimators initialized with any partition of the same set of observations gives precisely the same results for all estimands (this is readily apparent from equations \eqref{eq:univariateUpdate}, \eqref{eq:bivariateUpdate}, \eqref{eq:univariatMergeNonStd} and \eqref{eq:bivariateMergeNonStd}). Merging is not as efficient however, in that there is computational overhead in constructing the individual Hermite estimators to be merged. When observations are standardized, the resulting estimand values are expected to differ between a sequentially updated estimator and an estimator formed from merging however (although the differences are likely to be small). Conceptually, sequential updating of a Hermite series based estimator is best applied to incorporating new observations into the estimator in an efficient manner. Merging estimators on the other hand is perhaps most naturally applied in calculating marginal estimands. For example, if individual Hermite series based estimators are formed from observations for each hour of the day or for different sensors, merging the estimators for all hours or all sensors respectively corresponds to marginalizing the distribution of observations over these variables. We make these considerations more concrete in the numerical example in section \ref{sec:numerical_example}.

\section{Function reference and code usage examples} \label{sec:examples}

\subsection{Function reference summary} \label{subsec:func_ref}

We present a summary of the core functions provided by the \textsf{hermiter} package in Table \ref{tab:funcion_ref} and elaborate on the functions in more detail in the remainder of the section. 

\begin{table}
\caption{Summary of core functions provided by the \textsf{hermiter} package}
\begin{tabular}[t]{ll}
\hline
Function Name & Description\\
\hline
\textit{cum\_prob} & Estimate the cumulative probability at one or more x values \\
\textit{dens} & Estimate the probability density at one or more x values \\
\textit{density} & Return an object that can be printed and plotted using \\
& the generic methods \textit{print} and \textit{plot} respectively (adds a\\
& further generic density method to \textbf{R})\\
\textit{hcdf} & Return an object that can be printed, plotted and summarized \\
& using the generic methods \textit{print}, \textit{plot} and \textit{summary} respectively  \\
\textit{hermite\_estimator} & Construct a univariate or bivariate \texttt{hermite\_estimator} object\\
\textit{kendall} & Estimate the Kendall rank correlation coefficient for a bivariate\\ 
& estimator \\
\textit{merge\_hermite} & Merge a list of Hermite estimators\\
\textit{quant} & Estimate the quantiles at a vector of probability values for \\
& a univarite estimator\\
\textit{quantile} & Dispatch to \textit{quant} method (adds a further generic quantile\\
& method to \textbf{R}) \\
\textit{spearmans} & Estimate the Spearman rank correlation coefficient for a bivariate\\
& estimator\\
\textit{update\_sequential} & Update a Hermite series based estimator sequentially \\
\hline
\end{tabular} 
\label{tab:funcion_ref}
\end{table}

\subsection{Constructing the estimator} \label{subsec:construct_est}

A \texttt{hermite\_estimator} \textbf{R} S3 object is constructed as below. The argument, \texttt{N}, adjusts the number of terms in the Hermite series based estimator and controls the trade-off between bias and variance. A lower \texttt{N} value implies a higher bias but lower variance and vice versa for higher values of \texttt{N}. We have found that empirically, different contexts  benefit from different default values of \texttt{N} for good performance. We have distilled our experience on a wide array of data sets into defaults that have been implemented in the \textsf{hermiter} package. For 
univariate, non-exponentially weighted estimators, the default is \texttt{N}$ = 50$ whereas for univariate, exponentially weighted estimators, the default is \texttt{N} $= 20$. For bivariate, non-exponentially weighted estimators, the default is \texttt{N} $= 30$.
Finally, for bivariate, exponentially weighted estimators, the default is \texttt{N} $= 20$. For further detail on selecting $N$ in the univariate setting, see section 4.1 of \cite{stephanou2017sequential} and for the bivariate setting, see section 3.3 of \cite{stephanou2021sequential}. The argument, \texttt{standardize}, controls whether or not to standardize observations before applying the estimator. Standardization usually yields better results and is recommended for most estimation settings, thus the default for \texttt{standardize} is set to \texttt{TRUE}. Finally, the estimator can be optionally initialized with a batch of observations via the argument \texttt{observations}. 

A univariate estimator is constructed as follows (the default estimator type is univariate, so this argument does not need to be explicitly set):
\begin{lstlisting}
hermite_est <- hermite_estimator(N=50, standardize=TRUE,
 est_type = "univariate")
\end{lstlisting}
Similarly for constructing a bivariate estimator:
\begin{lstlisting}
hermite_est <- hermite_estimator(N=30, standardize=TRUE, 
 est_type = "bivariate")
\end{lstlisting}
If one wishes to initialize the estimator with an initial batch of observations, the following syntax can be utilized. For univariate observations we have:
\begin{lstlisting}
hermite_est <- hermite_estimator(N=50, standardize=TRUE,
 observations = rlogis(n=1000))
\end{lstlisting}
For bivariate observations we have:
\begin{lstlisting}
hermite_est <- hermite_estimator(N=30, standardize=TRUE, 
 est_type = "bivariate", 
 observations = matrix(data = rnorm(2000),nrow = 1000,
 ncol=2))
\end{lstlisting}
Note that the constructors above only take numeric \texttt{observations} arguments i.e. a numeric vector or numeric matrix respectively. Initializing with data frames is not presently supported even if all variables are numeric.

\subsection{Sequential estimator updating in stationary scenarios} \label{subsec:seq_upd_stat}

In the sequential setting, observations are revealed one at a time. A 
\texttt{hermite\_estimator} object can be updated sequentially with a single or multiple new 
observations by utilizing the \textit{update\_sequential} method. Note that when updating the Hermite series based estimator sequentially, observations are also standardized sequentially if the \texttt{standardize} argument is set to \texttt{TRUE} in the constructor. This means that using the \textit{update\_sequential} method with \texttt{standardize} set to \texttt{TRUE} for an initial set of observations will give a slightly different result to initializing the estimator using the \texttt{observations} argument. For updating with univariate observations we have:
\begin{lstlisting}
observations <- rlogis(n=1000)
hermite_est <- hermite_estimator(N=50, standardize=TRUE)
hermite_est <- update_sequential(hermite_est,
	observations)
\end{lstlisting}
For bivariate observations we have:
\begin{lstlisting}
observations <- matrix(data = rnorm(2000),nrow = 1000, 
 ncol=2)
hermite_est <- hermite_estimator(N=30, standardize=TRUE, 
 est_type = "bivariate")
hermite_est <- update_sequential(hermite_est,
observations)
\end{lstlisting}
One can verify that the new observations have been incorporated by printing the \texttt{hermite\_estimator} object before and after the sequential updates: the number of observations can be seen to have increased by the appropriate amount. The generic method \textit{update\_sequential} dispatches to the appropriate implementation depending on whether the \texttt{hermite\_estimator} object is univariate or bivariate. In the univariate case, the sequential updating of the Hermite series coefficients is based on \eqref{eq:univariateUpdate}. In the bivariate case, the sequential updating of the Hermite series coefficients is based on \eqref{eq:bivariateUpdate}. 

\subsection{Sequential estimator updating in non-stationary scenarios} \label{subsec:seq_upd_non_stat} 

The \textsf{hermiter} package is also applicable to non-stationary data streams.
An exponentially weighted form of the Hermite series coefficients can be applied to handle 
this case. The estimators will adapt to the new distribution and incrementally “forget" the old distribution. In order to use the exponentially weighted form of the 
\texttt{hermite\_estimator}, the \texttt{exp\_weight\_lambda} argument must be set to a non-NA value.
Typical values for this parameter are 0.01, 0.05 and 0.1. The lower the 
exponential weighting parameter, the slower the estimator adapts and vice versa 
for higher values of the parameter. However, the variance of estimands obtained from the \texttt{hermite\_estimator} object increases with higher 
values of \texttt{exp\_weight\_lambda}, so there is a trade-off to bear in mind. For univariate observations we have:
\begin{lstlisting}
hermite_est <- hermite_estimator(N=20, standardize=TRUE,
 exp_weight_lambda = 0.01)
for (idx in seq_len(1000)) {
	observation <- rnorm(1,mean=0.001*idx)
	hermite_est <-
	 update_sequential(hermite_est,observation)
}
\end{lstlisting}
For bivariate observations we have:
\begin{lstlisting}
hermite_est <- hermite_estimator(N=20, standardize=TRUE, 
 exp_weight_lambda = 0.01, est_type = "bivariate")
for (idx in seq_len(1000)) {
	observation <- rnorm(2,mean=0.001*idx)
	hermite_est <- update_sequential(hermite_est,
	 observation)
}
\end{lstlisting}
Again the generic method \textit{update\_sequential} dispatches to the appropriate implementation depending on whether the \texttt{hermite\_estimator} object is univariate or bivariate. In the univariate case, the sequential updating of the Hermite series coefficients is based on \eqref{eq:univariateUpdateExp}. In the bivariate case, the sequential updating of the Hermite series coefficients is based on \eqref{eq:bivariateUpdateExp}. 

\subsection{Merging Hermite estimators} \label{subsec:merge_est}

Hermite series based estimators can be consistently merged in both the univariate and bivariate settings as discussed previously. In particular, when \texttt{standardize} = \texttt{FALSE}, the results obtained from merging distinct \texttt{hermite\_estimator} objects updated on subsets of a data set are exactly equal to those obtained by constructing a single \texttt{hermite\_estimator} and updating on the full data set (corresponding to the concatenation of the aforementioned subsets). This holds true for the PDF, CDF and quantile results in the univariate case and the PDF, CDF and nonparametric correlation results in the bivariate case. When \texttt{standardize} = \texttt{TRUE}, the equivalence is no longer exact, but is accurate enough to be practically useful. Merging \texttt{hermite\_estimator} objects is illustrated below. Note that the \textit{merge\_hermite} function takes in a list of \texttt{hermite\_estimator} objects as an argument. In order to be merged, these objects must have a consistent type, either \texttt{"univariate"} or \texttt{"bivariate"}. As an example in the univariate case:
\begin{lstlisting}
set.seed(10)
observations_1 <- rlogis(n=1000)
observations_2 <- rlogis(n=1000)
hermite_est_1 <- hermite_estimator(N=50,standardize=FALSE,
 observations = observations_1)
hermite_est_2 <- hermite_estimator(N=50,standardize=FALSE,
 observations = observations_2)
hermite_est_merged <- merge_hermite(list(hermite_est_1, 
hermite_est_2))
hermite_est_full  <- hermite_estimator(N=50,standardize=FALSE,observations=c(observations_1,observations_2))
all.equal(hermite_est_merged,hermite_est_full)
\end{lstlisting}
The above example yields the following output which verifies that the merged \texttt{hermite\_estimator} i.e. \texttt{hermite\_est\_merged} is exactly equal equal to the full \texttt{hermite\_estimator}, i.e. \texttt{hermite\_est\_full} when \texttt{standardize} = \texttt{FALSE}.
\begin{lstlisting}
> all.equal(hermite_est_merged,hermite_est_full)
[1] TRUE
\end{lstlisting}
A univariate example with \texttt{standardize} = \texttt{TRUE} is the following:
\begin{lstlisting}
observations_1 <- rlogis(n=1000)
observations_2 <- rlogis(n=1000)
hermite_est_1 <- hermite_estimator(N=50,standardize=TRUE,observations=observations_1)
hermite_est_2 <- hermite_estimator(N=50,standardize=TRUE,observations=observations_2)
hermite_est_merged <- merge_hermite(list(hermite_est_1,hermite_est_2))
hermite_est_full  <- hermite_estimator(N=50,standardize=TRUE,observations=c(observations_1,observations_2))
all.equal(hermite_est_merged,hermite_est_full)
\end{lstlisting}
This yields the output:
\begin{lstlisting}
> all.equal(hermite_est_merged,hermite_est_full)
[1] "Component `coeff_vec': Mean relative difference: 0.006074491"
\end{lstlisting}
The above example illustrates that the merged \texttt{hermite\_estimator} i.e. \texttt{hermite\_est\_merged} is a close approximation to the full \texttt{hermite\_estimator}, i.e. \texttt{hermite\_est\_full} when \texttt{standardize} = \texttt{TRUE} but it is not exactly the same object. As an example for the bivariate case:
\begin{lstlisting}
observations_1 <- matrix(data = rnorm(2000), 
nrow = 1000, ncol=2)
observations_2 <- matrix(data = rnorm(2000), 
nrow = 1000, ncol=2)
hermite_est_1 <- hermite_estimator(N=30,standardize=TRUE, 
 est_type = "bivariate", observations = observations_1)
hermite_est_2 <- hermite_estimator(N=30,standardize=TRUE, 
 est_type = "bivariate", observations = observations_2) 
hermite_est_merged <- merge_hermite(list(hermite_est_1, 
hermite_est_2))
\end{lstlisting}
When \texttt{standardize} = \texttt{FALSE}, merging utilizes \eqref{eq:univariatMergeNonStd} and \eqref{eq:bivariateMergeNonStd} to combine the univariate and bivariate Hermite series coefficients respectively. When \texttt{standardize} = \texttt{TRUE}, merging utilizes the methods in \cite{schubert2018numerically} to combine the mean and standard deviation estimates of the individual \texttt{hermite\_estimator} objects along with \eqref{eq:univariatMergeStd} and \eqref{eq:bivariatMergeStd} to combine the univariate and bivariate Hermite series coefficients respectively.

A few final notes are that the \texttt{standardize} parameter must be the same i.e. either all \texttt{TRUE} or all \texttt{FALSE} for the list of \texttt{hermite\_estimator} objects to be merged. In addition, the \texttt{N} argument must be the same for all estimators in order to merge them.

\subsection{Estimating the univariate PDF, CDF and quantile function} \label{subsec:univariate_est}

The main advantage of Hermite series based estimators is that they can be updated in a sequential/one-pass manner as above and subsequently probability densities and cumulative probabilities at arbitrary x values can be obtained, along with arbitrary quantiles. The \texttt{hermite\_estimator} object only maintains a small and fixed number of coefficients and thus uses minimal memory. The syntax to calculate probability densities, cumulative probabilities and quantiles in the univariate setting is presented below. The PDF estimators \eqref{eq:HermiteSeriesProbEst} and \eqref{eq:HermiteSeriesProbEstBivariate} are implemented in the \textit{dens} method, where dispatch occurs to the appropriate implementation depending on whether the \texttt{hermite\_estimator} object is univariate or bivariate. As described in section \ref{sec:limitations}, density estimates can be negative in principle. The parameter \texttt{clipped} can be set to \texttt{TRUE} in order to replace negative values with a small, but strictly positive value (i.e. 1e-8). 

The CDF estimators \eqref{eq:HermiteSeriesUniCDFEst} and \eqref{eq:HermiteSeriesCDFEstBivariate} are implemented in the  \textit{cum\_prob} method, where again dispatch occurs to the appropriate implementation depending on whether the \texttt{hermite\_estimator} object is univariate or bivariate. Cumulative probability estimates may not be monotonically non-decreasing and might lie outside the range $[0,1]$ as per section \ref{sec:limitations}. These limitations can be partially addressed by clipping the estimates to lie between $[0,1]$, this can be done by setting the \texttt{clipped} parameter to \texttt{TRUE}. There are also various post-processing strategies that can be employed to enforce the monotone property (e.g. making use of the \textbf{R} function \textit{cummax}). The quantile function \textit{quant} makes use of the CDF estimator \eqref{eq:HermiteSeriesUniCDFEstAlt} and is only defined for univariate \texttt{hermite\_estimator} objects. It is worth noting that since the Hermite series based quantile estimator is directly estimating the quantile function and not the empirical quantiles, we do not expect the 0-quantile to correspond to the minimum observation, nor do we expect the 1-quantile to correspond to the maximum observation. In fact, since the Hermite series based estimators are defined on the full real line, the 0-quantile (true minimum) is $-\infty$ and the 1-quantile (true maximum) is $\infty$ by definition. As a result, estimating these quantiles, i.e. estimating the minimum and maximum directly is not sensible using these estimators. Quantiles for $p$ very close to 0 and $p$ very close to 1 may still be reasonably accurate however. The default behaviour of the \textit{quant} function is therefore to return a quantile for $p$ very close to 0 for the 0-quantile and a quantile for $p$ very close to 1 for the 1-quantile.

\begin{figure}[t!]
\centering
\includegraphics[width=80mm]{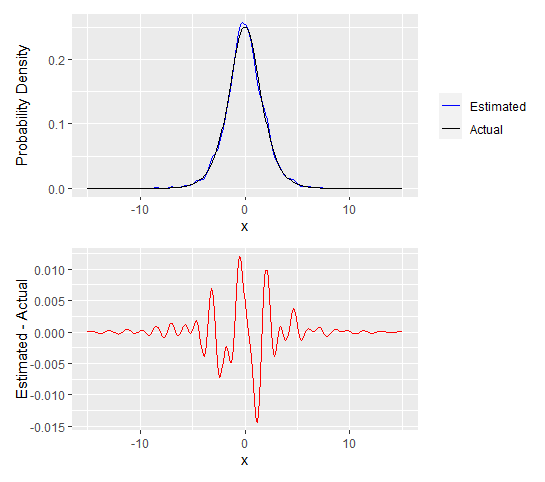}
\caption{\label{fig:pdf_univariate} Estimated (blue) versus actual (black) PDF. The difference between estimated and actual density values is plotted in red.}
\end{figure}

\begin{figure}[t!]
\centering
\includegraphics[width=80mm]{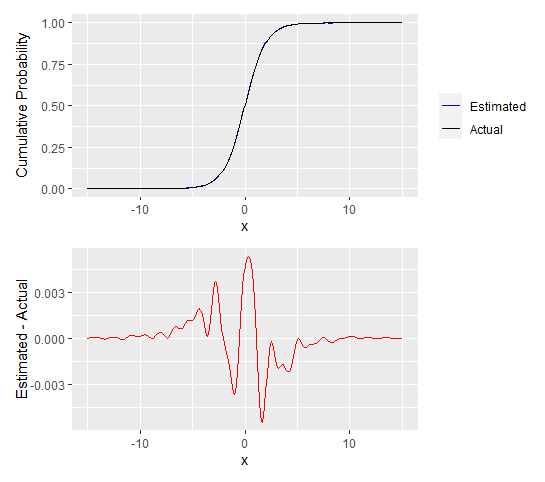}
\caption{\label{fig:cdf_univariate} Estimated (blue) versus actual (black) CDF. The difference between estimated and actual cumulative probability values is plotted in red.}
\end{figure}

\begin{figure}[t!]
\centering
\includegraphics[width=80mm]{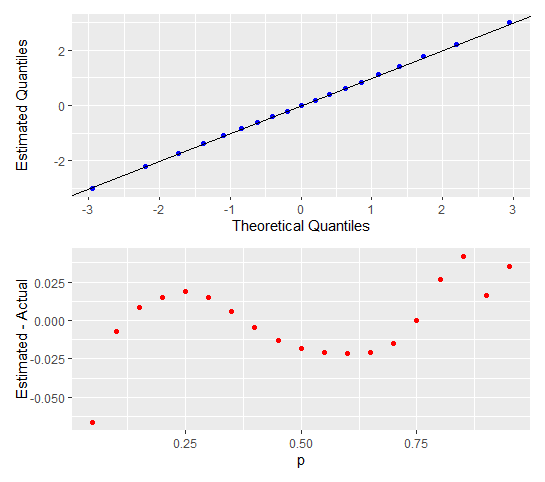}
\caption{\label{fig:quantiles} Q-Q Plot of estimated versus actual quantiles. The difference between estimated and actual quantile values is plotted in red.}
\end{figure}

As an example of estimating the PDF, CDF and quantiles in the univariate case: \\
\begin{lstlisting}
set.seed(10)
hermite_est <- hermite_estimator(N=50, standardize=TRUE,
 observations = rlogis(n=5000))
x <- seq(-15,15,0.1)
pdf_est <- dens(hermite_est,x)
cdf_est <- cum_prob(hermite_est,x)
p <- seq(0.05,0.95,0.05)
quantile_est <- quant(hermite_est,p)
\end{lstlisting}
Actual values of the PDF, CDF and quantiles are evaluated as below:
\begin{lstlisting}
actual_pdf <- dlogis(x)
actual_cdf <- plogis(x)
df_pdf_cdf <- data.frame(x,pdf_est,cdf_est,actual_pdf,
 actual_cdf)
actual_quantiles <- qlogis(p)
df_quant <- data.frame(p,quantile_est,actual_quantiles)
\end{lstlisting}
The estimated versus actual values are visualized in Figures \ref{fig:pdf_univariate}, \ref{fig:cdf_univariate} and \ref{fig:quantiles}.

\newpage
\subsection{Estimating the bivariate PDF, CDF and nonparametric correlations} \label{subsec:bivariate_est}

\begin{figure}[t!]
\centering
\includegraphics[width=95mm]{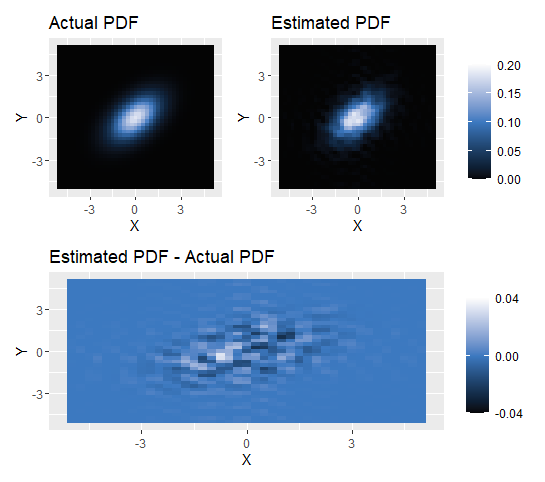}
\caption{\label{fig:pdf_bivariate} Actual bivariate PDF (left) versus estimated bivariate PDF (right).}
\end{figure}

\begin{figure}[t!]
\centering
\includegraphics[width=95mm]{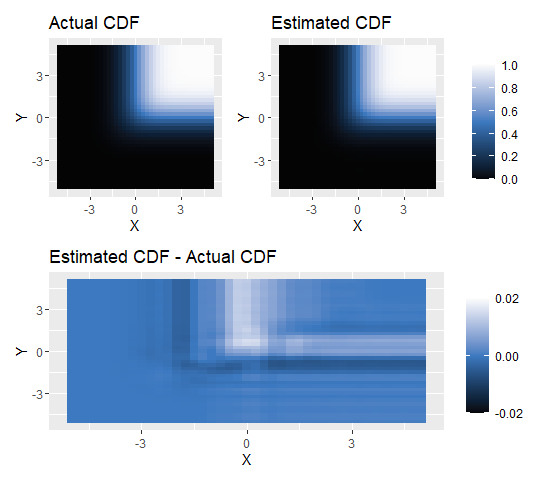}
\caption{\label{fig:cdf_bivariate} Actual bivariate CDF (left) versus estimated bivariate CDF (right).}
\end{figure}

The aforementioned suitability of Hermite series based estimators in sequential and one-pass batch estimation settings extends to the bivariate case. Probability densities and cumulative probabilities can be obtained at arbitrary points. The syntax to calculate probability densities and cumulative probabilities along with the Spearman and Kendall correlation coefficients in the bivariate setting is presented below. Note that the functions \textit{spearmans} and \textit{kendall} implement the equations \eqref{eq:hermiteSpearmansEstMat} and \eqref{eq:hermiteKendallEstMat} respectively. As an example of estimating the PDF, CDF and nonparametric correlation coefficients in the bivariate case:
\begin{lstlisting}
sig_x <- 1
sig_y <- 1
num_obs <- 5000
rho <- 0.5
observations_mat <- mvtnorm::rmvnorm(n=num_obs, 
 mean= rep(0,2),sigma = matrix(c(sig_x^2,rho*sig_x*sig_y,
 rho*sig_x*sig_y, sig_y^2), nrow=2, ncol=2, byrow= TRUE))
hermite_est <- hermite_estimator(N = 30, 
 standardize = TRUE, est_type = "bivariate", 
 observations = observations_mat)
vals <- seq(-5,5,by=0.25)
x_grid <- as.matrix(expand.grid(X=vals, Y=vals))
pdf_est <- dens(hermite_est, x_grid, clipped = TRUE)
cdf_est <- cum_prob(hermite_est, x_grid, clipped = TRUE)
spear_est <- spearmans(hermite_est)
kendall_est <- kendall(hermite_est)
\end{lstlisting}
Actual values of the bivariate PDF, CDF, Spearman and Kendall correlation coefficients are evaluated as below.
\begin{lstlisting}
actual_pdf <-mvtnorm::dmvnorm(x_grid,mean=rep(0,2), 
 sigma = matrix(c(sig_x^2,rho*sig_x*sig_y,
 rho*sig_x*sig_y,sig_y^2), nrow=2, ncol=2, byrow = TRUE))
actual_cdf <- rep(NA,nrow(x_grid))
for (row_idx in seq_len(nrow(x_grid))) {
	actual_cdf[row_idx] <-  
	 mvtnorm::pmvnorm(lower = c(-Inf, -Inf), 
	 upper=as.numeric(x_grid[row_idx,]), 
	 mean=rep(0,2),sigma = matrix(c(sig_x^2, 
	 rho*sig_x*sig_y,rho*sig_x*sig_y,sig_y^2), 
	 nrow=2, ncol=2, byrow = TRUE))
}
actual_spearmans <- cor(observations_mat, 
 method = "spearman")[1,2]
actual_kendall <- cor(observations_mat,
 method = "kendall")[1,2]
df_pdf_cdf <- data.frame(x_grid,pdf_est,cdf_est, 
actual_pdf,actual_cdf)
\end{lstlisting}
The estimated versus actual values for the PDF and CDF are visualized in Figures \ref{fig:pdf_bivariate}, \ref{fig:cdf_bivariate} respectively. 
The Spearman and Kendall correlation coefficient results are presented in Table \ref{tab:spearAndKendallResults}.
\begin{table}[t!]
\centering
\caption{Estimated versus actual Spearman Rho and Kendall Tau.}
{\small
\begin{tabular}{lll}
\hline
 & Spearman Rho & Kendall Tau\\
\hline
Actual & 0.479 & 0.331\\
Estimated & 0.475 & 0.330\\
\hline
\end{tabular}
\label{tab:spearAndKendallResults}
}
\end{table}

\newpage
\section{Numerical example} \label{sec:numerical_example}

In this section, we illustrate the utility of the \textsf{hermiter} package on a real data set.  The data set consists of one week (the first week of October 2021) of high frequency forex quote data sourced from TrueFX \citep{truefx} for the currency pairs EUR/USD and GBP/USD. These data are comprised of tick-by-tick, top-of-book bid and offer quotes, aggregated across several bank, broker and asset manager participants on the Integral OCX platform\footnote{This platform is an ECN i.e. an Electronic Communication Network}. In particular, we use the \textsf{hermiter} package to summarize the quantiles of the bid-offer spreads of EUR/USD and GBP/USD per hour of day, and merge these estimates to obtain the quantiles of spread over all hours (i.e. we marginalize the distribution of spreads over hour of day). This is contrasted with tracking the spread quantiles in a sequential manner. Finally we illustrate tracking the Spearman and Kendall correlation coefficients of the spreads of the EUR/USD and GBP/USD currency pairs in a sequential manner.  Bid-offer spread is reflective of the cost to liquidity takers of fulfilling immediate liquidity needs i.e. the need to buy or sell at any given time, and is defined as the prevailing offer price minus the prevailing bid price. We sample the spread at 15 second intervals. 

The quantile results are obtained by creating a univariate \texttt{hermite\_estimator} object for each hour of the day (each initialized with a batch of spread observations corresponding to a particular hour) and running the \textit{quant} function for $p=0.01, 0.1, 0.25, 0.5, 0.75, 0.9, 0.99$. The \textit{merge\_hermite} function is then used to combine the estimates (marginalize the distribution of spreads over hour of day) and the \textit{quant} function is again applied to the merged estimator. The results are presented in Table \ref{tab:currency_pair_quants}.  This procedure illustrates the utility of fast batch calculation of quantiles for each hour of day and then merging the estimators to marginalize the distribution over all hours of day and obtain combined quantile estimates. It is noteworthy that estimates can just as easily be obtained for any subset of hours of the day, avoiding the need to cater for a combinatorial explosion of combinations upfront. A very practical application is to calculate statistics for different trading sessions for example e.g. Tokyo, London and New York. The results suggest a mostly stable median spread throughout the day for both EUR/USD and GBP/USD, which is consistent with the fact that these are two of the most liquid currency pairs and are heavily traded in all trading sessions (covering almost 24 hours). The exceptions to this are the hours 21h and 22h i.e. 9 pm and 10 pm UTC time. These hours coincide with a period where trading in New York tapers off and trading in Tokyo has not yet picked up. These hours are thus typically associated with the lowest traded volumes of the day and this is reflected in the median spreads in Table \ref{tab:currency_pair_quants} which are higher than the rest of the day.

\begin{figure}[t!]
\centering
\includegraphics[width=75mm]{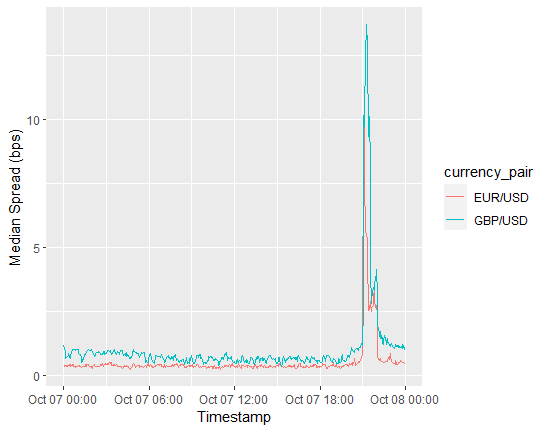}
\caption{\label{fig:dynamic_spread_quantiles} Sequentially updated median estimate of spread values for EUR/USD and GBP/USD.}
\end{figure}

\begin{figure}[t]
\centering
\includegraphics[width=75mm]{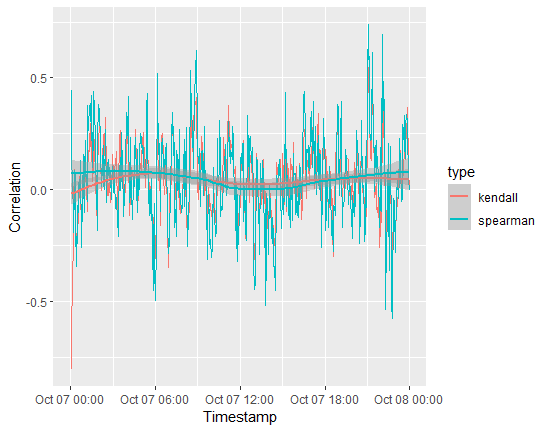}
\caption{\label{fig:dynamic_spread_corr} Sequentially updated estimates of the Spearman and Kendall correlations of spread values for EUR/USD and GBP/USD.}
\end{figure}

A distinct, but related use case of  \textsf{hermiter} is to track time-varying quantiles in a sequential manner. In this case, the exponentially weighted form of the univariate Hermite series based estimator can be applied (where the \texttt{hermite\_estimator} object is contructed with a non-NA weighting parameter $\lambda$). In the present numerical example, a \texttt{hermite\_estimator} object was contructed with $\lambda=0.01$, and the estimator was sequentially updated with one spread observation at a time using the \textit{update\_sequential} function. The updated median value was calculated on each update. The result of this procedure is illustrated in Figure \ref{fig:dynamic_spread_quantiles}. This figure again reflects the spike in median spread between 9 pm and 11 pm.  We illustrate a similar procedure of tracking time-varying nonparametric correlations between EUR/USD and GBP/USD spreads using a bivariate \texttt{hermite\_estimator} and the \textit{spearmans}, \textit{kendall} functions, calculated at each update, in Figure \ref{fig:dynamic_spread_corr}. There are several interesting features of this figure. The first is that the correlation between the currency pair spreads varies significantly throughout the day. The second interesting feature is that the local mean levels of the correlations are very slightly positive throughout the day. Lastly, there are periods where the correlation spikes above 0.5 in magnitude, which is meaningful information for risk management. A final note is that we found it advantageous to apply a log transformation to the positive spread data before constructing \texttt{hermite\_estimator} objects in this section (we invert the transformation for the results) as this yielded more sensible summary statistics.

\begin{table}[h!]
\caption{Quantile estimates for spreads of EUR/USD and GBP/USD in basis points per hour of day and overall (where the distribution has been marginalized over hour of day).}
{\tiny
\begin{tabular}{llrrrrrrr}
\toprule
Currency Pair & Hour (UTC) & p\_1\% & p\_10\% & p\_25\% & p\_50\% & p\_75\% & p\_90\% & p\_99\%\\
\midrule
EUR/USD & 0 & 0.2 & 0.3 & 0.3 & 0.4 & 0.4 & 0.5 & 0.7\\
EUR/USD & 1 & 0.2 & 0.3 & 0.3 & 0.4 & 0.4 & 0.4 & 0.5\\
EUR/USD & 2 & 0.2 & 0.3 & 0.3 & 0.4 & 0.4 & 0.5 & 0.6\\
EUR/USD & 3 & 0.2 & 0.3 & 0.3 & 0.4 & 0.4 & 0.5 & 0.6\\
EUR/USD & 4 & 0.2 & 0.3 & 0.3 & 0.4 & 0.4 & 0.5 & 0.6\\
EUR/USD & 5 & 0.2 & 0.3 & 0.3 & 0.4 & 0.4 & 0.5 & 0.6\\
EUR/USD & 6 & 0.2 & 0.3 & 0.3 & 0.4 & 0.4 & 0.4 & 0.5\\
EUR/USD & 7 & 0.2 & 0.3 & 0.3 & 0.4 & 0.4 & 0.5 & 0.5\\
EUR/USD & 8 & 0.2 & 0.3 & 0.3 & 0.4 & 0.4 & 0.4 & 0.5\\
EUR/USD & 9 & 0.2 & 0.3 & 0.3 & 0.4 & 0.4 & 0.4 & 0.5\\
EUR/USD & 10 & 0.1 & 0.2 & 0.3 & 0.4 & 0.4 & 0.5 & 0.6\\
EUR/USD & 11 & 0.2 & 0.3 & 0.3 & 0.4 & 0.4 & 0.5 & 0.6\\
EUR/USD & 12 & 0.2 & 0.3 & 0.3 & 0.3 & 0.4 & 0.4 & 0.5\\
EUR/USD & 13 & 0.2 & 0.3 & 0.3 & 0.4 & 0.4 & 0.4 & 0.5\\
EUR/USD & 14 & 0.2 & 0.3 & 0.3 & 0.4 & 0.4 & 0.5 & 0.5\\
EUR/USD & 15 & 0.2 & 0.3 & 0.3 & 0.4 & 0.4 & 0.5 & 0.5\\
EUR/USD & 16 & 0.2 & 0.3 & 0.3 & 0.4 & 0.4 & 0.5 & 0.5\\
EUR/USD & 17 & 0.2 & 0.3 & 0.3 & 0.4 & 0.4 & 0.5 & 0.6\\
EUR/USD & 18 & 0.2 & 0.3 & 0.3 & 0.4 & 0.4 & 0.5 & 0.6\\
EUR/USD & 19 & 0.2 & 0.3 & 0.3 & 0.4 & 0.4 & 0.5 & 0.6\\
EUR/USD & 20 & 0.3 & 0.3 & 0.4 & 0.5 & 0.7 & 0.8 & 1.0\\
EUR/USD & 21 & 1.8 & 2.6 & 3.0 & 4.0 & 6.0 & 8.5 & 15.9\\
EUR/USD & 22 & 0.4 & 0.5 & 0.6 & 0.7 & 0.7 & 0.9 & 1.2\\
EUR/USD & 23 & 0.3 & 0.4 & 0.5 & 0.5 & 0.6 & 0.7 & 0.7\\
GBP/USD & 0 & 0.4 & 0.5 & 0.6 & 0.7 & 0.9 & 1.0 & 1.2\\
GBP/USD & 1 & 0.4 & 0.5 & 0.6 & 0.8 & 0.9 & 1.0 & 1.2\\
GBP/USD & 2 & 0.4 & 0.5 & 0.5 & 0.7 & 0.9 & 1.0 & 1.2\\
GBP/USD & 3 & 0.4 & 0.5 & 0.6 & 0.8 & 0.9 & 1.0 & 1.1\\
GBP/USD & 4 & 0.4 & 0.5 & 0.6 & 0.7 & 0.9 & 1.0 & 1.1\\
GBP/USD & 5 & 0.4 & 0.5 & 0.6 & 0.7 & 0.8 & 0.9 & 1.2\\
GBP/USD & 6 & 0.4 & 0.5 & 0.6 & 0.7 & 0.8 & 0.8 & 1.0\\
GBP/USD & 7 & 0.3 & 0.5 & 0.6 & 0.7 & 0.8 & 0.9 & 1.0\\
GBP/USD & 8 & 0.3 & 0.4 & 0.5 & 0.6 & 0.7 & 0.8 & 1.0\\
GBP/USD & 9 & 0.3 & 0.4 & 0.5 & 0.6 & 0.7 & 0.8 & 1.0\\
GBP/USD & 10 & 0.3 & 0.4 & 0.5 & 0.6 & 0.7 & 0.8 & 0.9\\
GBP/USD & 11 & 0.4 & 0.5 & 0.6 & 0.7 & 0.7 & 0.8 & 0.9\\
GBP/USD & 12 & 0.3 & 0.4 & 0.5 & 0.6 & 0.7 & 0.8 & 1.0\\
GBP/USD & 13 & 0.3 & 0.4 & 0.5 & 0.6 & 0.7 & 0.8 & 0.9\\
GBP/USD & 14 & 0.3 & 0.4 & 0.5 & 0.6 & 0.7 & 0.8 & 0.9\\
GBP/USD & 15 & 0.3 & 0.4 & 0.5 & 0.6 & 0.7 & 0.7 & 0.9\\
GBP/USD & 16 & 0.3 & 0.4 & 0.5 & 0.6 & 0.7 & 0.8 & 0.9\\
GBP/USD & 17 & 0.3 & 0.4 & 0.5 & 0.6 & 0.7 & 0.8 & 0.9\\
GBP/USD & 18 & 0.4 & 0.5 & 0.6 & 0.7 & 0.8 & 0.8 & 0.9\\
GBP/USD & 19 & 0.3 & 0.5 & 0.6 & 0.7 & 0.8 & 0.8 & 1.0\\
GBP/USD & 20 & 0.5 & 0.7 & 0.8 & 1.0 & 1.1 & 1.3 & 2.3\\
GBP/USD & 21 & 2.1 & 2.5 & 3.3 & 5.0 & 8.9 & 12.0 & 24.7\\
GBP/USD & 22 & 0.9 & 1.0 & 1.1 & 1.3 & 1.6 & 1.9 & 2.5\\
GBP/USD & 23 & 0.7 & 0.8 & 0.9 & 1.0 & 1.2 & 1.4 & 1.6\\
EUR/USD & All & 0.2 & 0.3 & 0.3 & 0.4 & 0.5 & 0.6 & 6.4\\
GBP/USD & All & 0.4 & 0.5 & 0.6 & 0.7 & 0.9 & 1.1 & 9.0\\
\bottomrule
\end{tabular}
\label{tab:currency_pair_quants}
}
\end{table}

\section{Accuracy and performance} \label{sec:acc_perf}

\subsection{Univariate setting}\label{subsec:acc_perf_univariate}

In the univariate case, simulation studies have been conducted to demonstrate competitive performance of the Hermite series based algorithms in the sequential quantile estimation setting in \cite{stephanou2017sequential}. In addition, simulation studies have been performed to evaluate the accuracy of the Hermite series based CDF estimator in \cite{stephanou2021properties} where it was concluded that while smooth kernel distribution function estimators are more accurate in the non-sequential setting, the performance of the sequential Hermite series based CDF estimators is still sufficiently good to be practically useful in online settings. The recent \textsf{tdigest} package in \textbf{R} offers very similar functionality for sequential quantile estimation to \textsf{hermiter}. This similarity, along with the availability of a \textbf{R} implementation and the popularity of the tdigest algorithm suggests a new and useful comparison to the Hermite series based algorithms in terms of accuracy and performance in the stationary setting. Note that in the non-stationary setting, where quantiles may vary over time, only the \textsf{hermiter} package can be directly applied, implying that even if the accuracy and performance of \textsf{tdigest} and \textsf{hermiter} are similar in the stationary setting, the \textsf{hermiter} package still adds valuable functionality.
The simulation comparison we conduct utilizes the suite of test distributions provided by the \textbf{R} package \textsf{benchden} \citep{mildenberger2012benchden} excluding heavy-tailed distributions. The reason we exclude the heavy-tailed densities from the \textsf{benchden} test suite is that most of these have undefined variances and some even have undefined means. The Hermite series based quantile estimators as implemented in \textsf{hermiter} require at least the first two moments to exist and thus we cannot meaningfully compare \textsf{hermiter} and \textsf{tdigest} on these densities. The simulation study is as follows: for sample sizes, $n=10,000; 100,000; 1000,000; 10,000,000$ and each of the test distributions, the following steps are repeated.

\begin{enumerate}
	\item For each test distribution, we initialize a \texttt{hermite\_estimator} object (using defaults \texttt{N} $=50$, \texttt{standardize} = \texttt{TRUE}), with $n$ observations drawn from the test distribution. We record the integrated absolute error (IAE) and partial integrated absolute error (pIAE) between estimated quantiles - obtained via the \textit{quant} function with \texttt{algorithm} = \texttt{"interpolate"} and \texttt{accelerate\_series} = \texttt{TRUE} - and the true quantile values. Similarly for the tdigest algorithm using the \textit{tdigest} and \textit{quantile} functions. The IAE is defined as,
	\begin{equation*}
		\mbox{IAE} = \int_{0}^{1} \lvert \hat{q}(p) - q(p)\rvert dp,
	\end{equation*}
	and the partial IAE is defined as,
		\begin{equation*}
		\mbox{pIAE} = \int_{0.01}^{0.99} \lvert \hat{q}(p) - q(p)\rvert dp,
		\end{equation*}
where $\hat{q}(p)$ is the estimated quantile value at cumulative probability $p$ and $q(p)$ is the true quantile value at cumulative probability $p$. The IAE captures the integrated absolute error across the full range of cumulative probability values. The partial IAE captures the integrated absolute error excluding extreme quantiles (namely those where $p<0.01$ and $p>0.99$). In practice, we apply Quasi-Monte Carlo integration with a Sobol sequence to numerically determine the value of these integrals.
	\item We repeat the above $m=100$ times and average across all runs to obtain estimates of the mean integrated absolute error, $\mbox{MIAE}=E\left(\mbox{IAE}\right)$, and partial mean integrated absolute error, $\mbox{pMIAE}=E\left(\mbox{pIAE}\right)$.
\end{enumerate}
\begin{table}[t!]
\caption{Comparison of \textsf{hermiter} and \textsf{tdigest} for different sample sizes. We report the fraction of distributions in the \textsf{benchden} test suite where the Hermite series based estimator has a lower mean integrated absolute error (MIAE) than the tdigest based algorithm. Fractions of greater than 0.5 are reported in bold.}
\centering
{\small
\begin{tabular}{rrr}
\toprule
Observation Count & Hermite Estimator Better MIAE\\
\midrule
10,000 & \textbf{0.57 (12/21)}\\
100,000 & \textbf{0.81 (17/21)}\\
1000,000 & \textbf{0.86 (18/21)}\\
10,000,000 & \textbf{0.86 (18/21)}\\
\bottomrule
\end{tabular}
\label{tab:hermitevtdigest}
}
\end{table}
\begin{table}[t!]
\caption{Comparison of \textsf{hermiter} and \textsf{tdigest} for different sample sizes. We report the fraction of distributions in the \textsf{benchden} test suite where the Hermite series based estimator has a lower partial mean integrated absolute error (pMIAE) than the tdigest based algorithm. Fractions of greater than 0.5 are reported in bold.}
\centering
{\small
\begin{tabular}{rrr}
\toprule
Observation Count & Hermite Estimator Better pMIAE\\
\midrule
10,000 & \textbf{0.62 (13/21)}\\
100,000 & \textbf{0.81 (17/21)}\\
1000,000 & \textbf{0.86 (18/21)}\\
10,000,000 & \textbf{0.91 (19/21)}\\
\bottomrule
\end{tabular}
\label{tab:hermitevtdigestp}
}
\end{table}
This yields a MIAE and pMIAE value for each test distribution and sample size $n$ for both the Hermite series based quantile estimator and tdigest based quantile estimation algorithm. There are $21$ non heavy-tailed test distributions in total. We summarize the results of comparing the MIAE of \textsf{hermiter} and \textsf{tdigest} estimates in Table \ref{tab:hermitevtdigest}. The results for pMIAE are presented in Table \ref{tab:hermitevtdigestp}.

The results reveal that the accuracy of the Hermite series based approach is better than the tdigest algorithm for fairly large numbers of observations even for the full MIAE error measure. In this simulation study we are particularly interested in performance on a number of common distributions across a range of non-extreme (i.e. more central) quantiles, for which the pMIAE results are more relevant. These results demonstrate significantly better accuracy using the Hermite series based algorithm for quantile estimates formed from more than $n=10,000$ observations. The distributions in \textsf{benchden} include those with full real-line, half real-line and compact support and thus the simulation study should yield information about typical performance in a broad range of scenarios. We reiterate that we exclude the heavy-tailed densities from the \textsf{benchden} test suite but should mention that the \textsf{tdigest} package appears to perform well in quantile estimation on all these heavy-tailed densities. Thus, for applications involving heavy-tailed distributions, \textsf{tdigest} is likely better suited. 

\begin{figure}[t!]
\centering
\includegraphics[width=70mm]{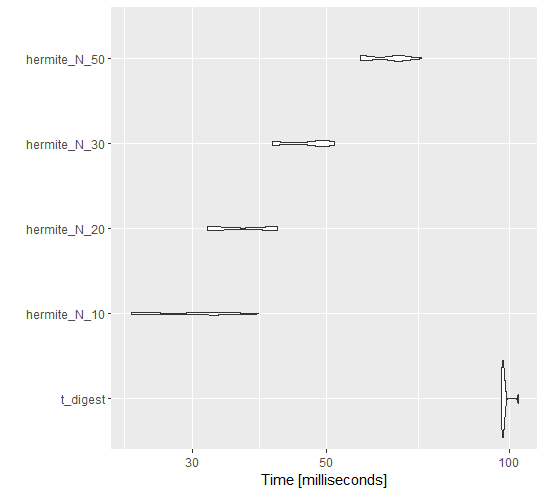}
\caption{\label{fig:t_digest_v_hermite} Computational performance comparison of functions in \textsf{tdigest} versus functions in \textsf{hermiter} for batch updates of $n=1000,000$ observations with \texttt{options(hermiter.parallel = TRUE)}.}
\end{figure}

\begin{figure}[t!]
\centering
\includegraphics[width=70mm]{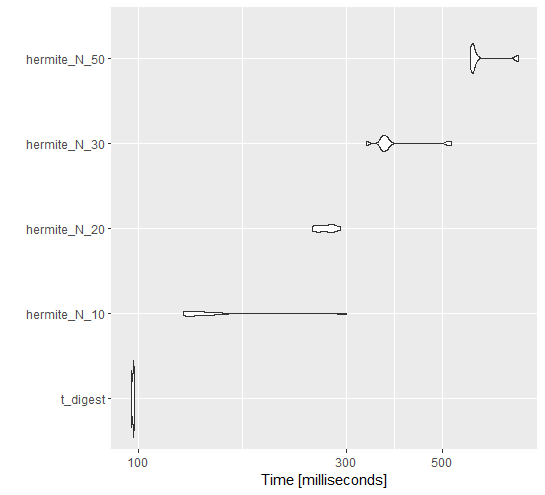}
\caption{\label{fig:t_digest_v_hermite_serial} Computational performance comparison of functions in \textsf{tdigest} versus functions in \textsf{hermiter} for batch updates of $n=1000,000$ observations with \texttt{options(hermiter.parallel = FALSE)}.}
\end{figure}

\begin{figure}[t!]
\centering
\includegraphics[width=75mm]{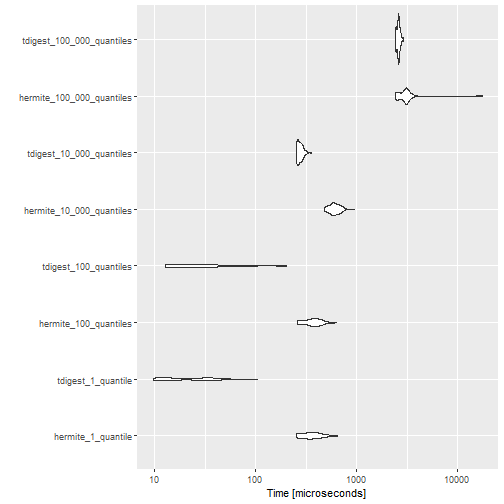}
\caption{\label{fig:t_digest_v_hermite_quant} Computational performance comparison of functions in \textsf{tdigest} versus functions in \textsf{hermiter} with \texttt{N} $=50$ for calculating 1 quantile, 100 quantiles, 10,000 quantiles and 100,000 quantiles. }
\end{figure}

In terms of computational performance, we plot the time taken to initialize the \textsf{hermiter} estimator and the \textsf{tdigest} estimator with $n=1000,000$ observations in Figure \ref{fig:t_digest_v_hermite}. It is apparent that \textsf{hermiter} is faster across all parameters of \texttt{N} considered. This is mainly the result of leveraging multithreaded computation on multiple cores for batch updating. This is enabled by default. Multithreaded computation can be disabled by setting \texttt{options(hermiter.parallel = FALSE)}. Computation time will typically increase on multi-core systems in this case however, compare Figure \ref{fig:t_digest_v_hermite} (multithreaded computation enabled) to Figure \ref{fig:t_digest_v_hermite_serial} (multithreaded computation disabled) for example. Both benchmarks were conducted on the same eight core machine (see section \ref{sec:comp_details}).
Finally, we plot the time taken to calculate vectors of quantile values ranging in length from a single quantile to 100,000 quantiles in Figure \ref{fig:t_digest_v_hermite_quant}. The performance of \textsf{tdigest} is better for smaller numbers of quantiles but the distributions of run-times of \textsf{hermiter} and \textsf{tdigest} are similar for larger numbers of quantiles. Calculating large vectors of quantiles simultaneously is meaningful for drawing random variables for example. Given the fact that further implementation optimization is possible for \textsf{hermiter} (and potentially \textsf{tdigest}) and that performance results are likely operating system dependent, we regard the computational performance as roughly similar (while acknowledging that \textsf{tdigest} appears faster for smaller vectors of quantiles, perhaps due to less computational overhead).

\subsection{Bivariate setting}\label{subsec:acc_perf_bivariate}

In the bivariate case, extensive simulation studies have been conducted to demonstrate competitive performance in the sequential estimation of the Spearman correlation coefficient versus the only known competing algorithm in \cite{stephanou2021sequential}. In this section we present a succinct simulation study using essentially the same methodology as \cite{stephanou2021sequential} where we extend the results to the sequential estimation of the Kendall Tau coefficient. We term the competing algorithms in \cite{xiao2019novel} count matrix based algorithms.

The simulation study involves generating i.i.d. samples of increasing size,  $n=10,000; 50,000; 100,000$, from a bivariate normal distribution with mean vector, $\mu = (0,0)$, and covariance matrix, $\Sigma = (\sigma_{1}, \rho \sigma_{1}\sigma_{2}; \rho\sigma_{1}\sigma_{2},\sigma_{2})$, where $\sigma_{1}=1, \sigma_{2}=1$. The correlation of the bivariate normal distribution is varied by drawing samples across a range of correlation parameters, $\rho = -0.75, -0.5, -0.25, 0.25,0.5, 0.75$. For each value of the sample size $n$ and correlation parameter, $\rho$, the below steps are repeated $m=100$ times.

\begin{enumerate}
\item Draw a sample of n i.i.d. observations, $(\mathbf{x_{i}},\mathbf{y_{i}}), \quad i=1,\dots,n$ from the bivariate normal distribution with correlation parameter $\rho$, mean vector $\mu$, and covariance matrix $\Sigma$, using the \textit{rmvnorm} function in the package \textsf{mvtnorm} \citep{mvtnormpkg,mvtnormBook}.
\item Construct a bivariate \texttt{hermite\_estimator} using \textsf{hermiter} and initialize with the full sample. 
\item Estimate the Spearman Rho and Kendall Tau correlation coefficients for the \\ \texttt{hermite\_estimator} object using the \textit{spearmans} and \textit{kendall} functions in \textsf{hermiter}.
\item Estimate the true Spearman Rho on the full sample using the \textbf{R} function \textit{cor}. Calculate the true Kendall Tau analytically using the relation $\tau=\frac{2}{\pi}\arcsin\rho$ which applies to bivariate normal distributions \citep{gibbons2010nonparametric}. We use the exact analytical Kendall Tau since the implementation in \textit{cor} is prohibitively slow for estimating Kendall Tau with larger sample sizes (as it would be in other standard implementations). 
\item Calculate the absolute error between the Hermite series based Spearman Rho and Kendall Tau estimates and the respective exact values.
\end{enumerate}

\begin{figure}[t!]
\centering
\includegraphics[width=55mm]{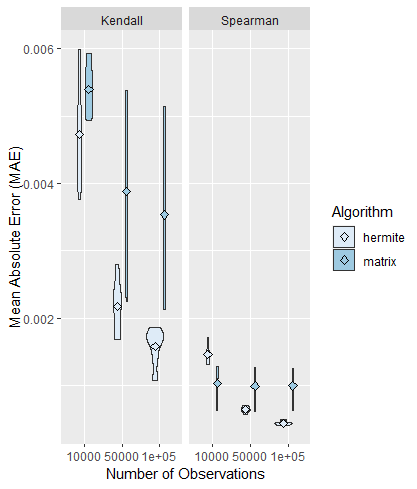}
\caption{\label{fig:mae_spearman_kendall}Violin plots of the MAE results for the Hermite series based algorithms and the count matrix based algorithms for Spearman Rho and Kendall Tau across all observation sizes. The mean MAE is represented by the diamond symbol for each set of values.}
\end{figure}

The mean absolute error (MAE) between the Hermite series based estimates and the exact Spearman and Kendall rank correlation coefficients for a particular value of $\rho$ and sample size $n$ are then estimated through $\mbox{MAE}(\hat{R}_{N}) = \frac{1}{m} \sum_{j=1}^{m} \lvert \hat{R}^{(j)}_{N} - R^{(j)} \rvert$ and $\mbox{MAE}(\hat{\tau}_{N}) = \frac{1}{m} \sum_{j=1}^{m} \vert \hat{\tau}^{(j)}_{N} - \tau^{(j)}\rvert$ respectively, where $j$ indexes a particular set of $n$ observations. The same steps as above are repeated for the count matrix algorithms namely, algorithms 2 and 3 of \cite{xiao2019novel}. For the choice of parameters, we use the default of \textsf{hermiter} for non-exponentially weighted, bivariate estimators, \texttt{N} $=30$, which we have found to give good performance in a wide range of scenarios. For algorithm 2 of \cite{xiao2019novel} we use the recommended value of 30 cutpoints for estimating Spearman Rho. For algorithm 3 of \cite{xiao2019novel} we use the recommended value of 100 cutpoints for estimating Kendall Tau. It is worth noting that the number of values to be maintained in memory is significantly larger in the count matrix case with 100 cutpoints than the Hermite estimator case with \texttt{N} $=30$. The results are summarized as the average MAE (and standard deviation of MAE) across all the values of $\rho$ in Table \ref{tab:spear_res} for the Hermite series based algorithm with \texttt{N} $=30$ and count matrix algorithm with $c=30$ for Spearman Rho estimation. Similarly, results are summarized as the average MAE (and standard deviation of MAE) across all the values of $\rho$ in Table \ref{tab:kendall_res} for the Hermite series based algorithm with \texttt{N} $=30$ and count matrix algorithm with $c=100$ for Kendall Tau estimation. It is clear that the Hermite series based algorithms present an accuracy advantage. The mean MAE and variation in MAE are also summarized in Figure \ref{fig:mae_spearman_kendall}.

The computational performance of the algorithms in \textsf{hermiter} is roughly comparable with the algorithms in \cite{xiao2019novel}. That said, when updating the Spearman or Kendall correlation coefficient with each new observation, we have observed better computational performance with the Hermite series based algorithms. We acknowledge that this may depend on the implementation of the count matrix algorithms however. We omit detailed benchmarking for brevity. A final note is that the calculation time for the Hermite series based Kendall Tau estimator is thousands of times faster than the \textit{cor} function in \textbf{R} with $n=10,000$ observations and this differential grows rapidly with $n$. For large samples (e.g. millions or even billions of observations), the standard methods of calculating the Kendall correlation coefficient are intractable. The Hermite series based approach is not only tractable but very fast in practice.
\begin{table}[t!]
\caption{A comparison of MAE results obtained for the Hermite series based Spearman correlation estimation algorithm (\texttt{N} $=30$) versus the count matrix based Spearman correlation estimation algorithm ($c = 30$). The lowest MAE values per sample size are reported in bold.}
{\footnotesize
\begin{tabular}{rrrrr}
\toprule
\multicolumn{1}{c}{ } & \multicolumn{2}{c}{Spearman Rho MAE Avg (x $10^{-2}$)} & \multicolumn{2}{c}{Spearman Rho MAE Std (x $10^{-2}$)} \\
\cmidrule(l{3pt}r{3pt}){2-3} \cmidrule(l{3pt}r{3pt}){4-5}
Observations & Matrix (c=30) & Hermite (N=30) & Matrix (c=30) & Hermite (N=30)\\
\midrule
10,000 & \textbf{0.103} & 0.146 & 0.029 & 0.014\\
50,000 & 0.099 & \textbf{0.065} & 0.028 & 0.005\\
100,000 & 0.100 & \textbf{0.044} & 0.029 & 0.003\\
\bottomrule
\end{tabular}
\label{tab:spear_res}
}
\end{table}
\begin{table}[t!]
\caption{A comparison of MAE results obtained for the Hermite series based Kendall correlation estimation algorithm (\texttt{N} $=30$) versus the count matrix based Kendall correlation estimation algorithm ($c = 100$). The lowest MAE values per sample size are reported in bold.}
{\footnotesize
\begin{tabular}{rrrrr}
\toprule
\multicolumn{1}{c}{ } & \multicolumn{2}{c}{Kendall Tau MAE Avg (x $10^{-2}$)} & \multicolumn{2}{c}{Kendall Tau MAE Std (x $10^{-2}$)} \\
\cmidrule(l{3pt}r{3pt}){2-3} \cmidrule(l{3pt}r{3pt}){4-5}
Observations & Matrix (c=100) & Hermite (N=30) & Matrix (c=100) & Hermite (N=30)\\
\midrule
10,000 & 0.540 & \textbf{0.472} & 0.038 & 0.077\\
50,000 & 0.388 & \textbf{0.217} & 0.127 & 0.042\\
100,000 & 0.353 & \textbf{0.158} & 0.134 & 0.029\\
\bottomrule
\end{tabular}
\label{tab:kendall_res}
}
\end{table}

\section{Summary and discussion} \label{sec:summary}

In this article we have introduced a new \textbf{R} package, \textsf{hermiter}, for sequential nonparametric estimation. The unifying theme of the package is the implementation of several recently developed algorithms for online sequential estimation based on Hermite series estimators. This package addresses several gaps in the \textbf{R} package ecosystem as well as the more general scientific computing ecosystem. Namely, the \textit{online} estimation of univariate PDFs, CDFs and quantiles, along with bivariate PDFs, CDFs and nonparametric correlation coefficients in both stationary \textit{and} non-stationary settings. We have demonstrated competitive performance versus leading existing approaches such as \textsf{tdigest} in \textbf{R} in the sequential quantile estimation setting. In addition, we have demonstrated improved accuracy in sequential nonparametric correlation estimation compared to the only alternate approach known to the authors. The \textsf{hermiter} package does not seek to be a comprehensive collection of online methods for statistical analysis in general, in contrast to a package such as \textsf{OnlineStats}. A work in progress by the authors is developing such a comprehensive online statistics package for \textbf{R} which draws on the functionality of \textsf{hermiter}. In summary, we expect \textsf{hermiter} to be a compelling choice for online estimation on streaming data and massive data sets.

%% -- Optional special unnumbered sections -------------------------------------

\section*{Computational details}\label{sec:comp_details}

The results in this paper were obtained using
\textbf{R}~4.3 \citep{rcore} running on Windows 10. The hardware utilized was an AMD Ryzen 4900H 8-core laptop with 64GB of RAM. The version of our \textbf{R} package utilized was \textsf{hermiter}~2.3.0 \citep{hermiterRpkg}. Versions of \textsf{hermiter} in development are available at \url{https://github.com/MikeJaredS/hermiter}. A script to facilitate reproduction of the results in this article is available at \url{https://github.com/MikeJaredS/article_hermiter}. The following packages were also utilized in preparation of this work, \textsf{benchden}~1.0.5 \citep{mildenberger2012benchden}, \textsf{BH}~1.78.0 \citep{bh}, \textsf{ggplot2}~3.4.0 \citep{ggplot2}, \textsf{microbenchmark}~1.4.9 \citep{microbenchmark}, \textsf{mvtnorm}~1.1-3 \citep{mvtnormpkg,mvtnormBook}, \textsf{patchwork}~1.1.2 \citep{patchwork}, \textsf{randtoolbox}~2.0.3 \citep{randtoolbox}, \textsf{Rcpp}~1.0.9 \citep{rcpp}, \textsf{RcppParallel}~5.1.5 \citep{rcppParallel}, \textsf{tdigest}~0.4.1 \citep{tdigestRpkg}. \textbf{R} and all packages used are available from the Comprehensive
\textbf{R} Archive Network (CRAN) at
\url{https://CRAN.R-project.org/}.

\bmhead{Statement and acknowledgements}

The views expressed in this article are those of the authors and do not necessarily reflect the views of Rand Merchant Bank. Rand Merchant Bank does not make any representations or give any warranties as to the correctness, accuracy or completeness of the information presented; nor does Rand Merchant Bank assume liability for any losses arising from errors or omissions in the information in this article. We would like to thank Ted Dunning for useful and interesting discussions. We would also like to sincerely thank the editor, associate editor and particularly the reviewers for thorough and deeply insightful feedback that helped us greatly improve this article.

\section*{Declarations}

On behalf of all authors, the corresponding author states that there is no conflict of interest.

\bibliography{article.bib}% common bib file

\end{document}